\documentclass[aps,prb,twocolumn,showpacs]{revtex4}
\usepackage{graphicx} 
\def\gapx{\lower 2pt \hbox{$\buildrel>\over{\scriptstyle{\sim}}$\ }}
\def\lapx{\lower 2pt \hbox{$\buildrel<\over{\scriptstyle{\sim}}$\ }}
\def\ph2{{\it p}-H$_2$}
\def\beq{\begin{equation}}
\def\eeq{\end{equation}}
\def\Am3{\AA$^{-3}$}
\begin{document}

\widetext
\title{Quantum statistics and the momentum distribution of liquid {\it para}-hydrogen}
\author{Massimo Boninsegni} 
\affiliation{Department of Physics, University of Alberta, Edmonton, Alberta, Canada T6G 2G7}
\date{\today}

\begin{abstract}
Extensive Monte Carlo simulations of bulk liquid {\it para}-hydrogen at a temperature $T$=16.5 K have been carried out using the continuous-space Worm Algorithm. Results for the momentum distribution, as well as for the kinetic energy per particle and the pair correlation function are reported.   The static equilibrium thermodynamic properties of this system can be generally computed by assuming that molecules are distinguishable. However, the one-body density matrix (and the associated momentum distribution) are affected by particle indistinguishability and quantum statistics, to an extent that lends itself to experimental observation.
Comparisons with available experimental data and other theoretical and numerical calculations are offered.
\end{abstract}
\maketitle
\section{Introduction}
The investigation of the condensed phases of molecular hydrogen is aimed at understanding and characterizing 
quantitatively quantum effects,  both in the liquid and in the solid phases. 
In many respects, liquid hydrogen displays a 
physical behavior that interpolates between that of a 
classical liquid, and that of superfluid helium, i.e., the most 
quantum-mechanical of the simple fluids.  
The most important manifestation of quantum mechanics in a fluid of hydrogen molecules, is the the fact that the kinetic energy per particle\cite{intro} significantly exceeds the classically predicted\cite{note0} value 3$T$/2. This is  essentially a consequence of zero-point motion, as the excess kinetic energy arises from the confinement that each molecule experiences inside the instantaneous ``cage" of surrounding particles, owing to the hard core of the inter-molecular interaction.

Hardly any signature of quantum statistics (Bose in this case, as both {\it para}- and {\it ortho}-hydrogen molecules have integer spin) can be detected on the equilibrium thermodynamic properties of liquid hydrogen, as particle exchanges are exceedingly rare. For a simple quantitative estimate, one may consider the lightest hydrogen isotope, i.e., {\it para}-hydrogen  (\ph2); under the pressure of its own vapor, liquid \ph2 crystallizes at a temperature $T$=13.8 K. At temperatures slightly above that, in the liquid phase, the thermal wavelength $\lambda_T$ of a \ph2 molecule is of the order of 1 \AA, significantly smaller than the mean intermolecular distance $d \sim$ 3.5 \AA. Thus, probability ``clouds" associated to different molecules do not overlap significantly, and exchanges are suppressed\cite{feynman} 
at least as $\sim$ exp$[-d^2/2\lambda_T^2] $, which is less than 10$^{-3}$ (for comparison, it is close to 0.3 in $^4$He at $T$=1 K). 

The presence of the above-mentioned hard core repulsion at distances less than $\sim$ 2.2 \AA\ in the interaction between two molecules, suppresses quantum exchanges even further.\cite{qbl} Consequently, liquid hydrogen has been modeled as a system of {\it distinguishable} quantum particles (``Boltzmannons"), e.g., in all numerical simulation work to date.\cite{neumann91,zoppib,wang97,kin,milva2,znc} Indeed, it is precisely the impossibility of stabilizing a phase of this system where quantum exchanges may be important,\cite{note} that has so far hampered the experimental observation of  superfluidity in bulk {\it para}-hydrogen. 

However, effects of particle indistinguishability and quantum statistics {\it ought} to be detectable in the {\it momentum distribution}, defined as 
\beq
\bar n({\bf k}) = \langle \hat a^\dagger_{\bf k}\hat a_{\bf k}\rangle
\eeq  
where $\langle ...\rangle$ stands for thermal expectation value, and $\hat a^\dagger_{\bf k}$ ($\hat a_{\bf k}$) is 
the creation (annihilation) operator for a particle of momentum {\bf k}. The momentum distribution is related, via a Fourier transformation, to the {\it one-body density matrix} \beq n({\bf r},{\bf r}^\prime)
= \langle \hat\psi^\dagger({\bf r})\ \hat\psi({\bf r}^\prime)\rangle\label {rel} \eeq
where $\hat\psi$ and $\hat\psi^\dagger$ are Bose field operators. Within the path integral formulation of quantum statistical mechanics,\cite{feynman} $n({\bf r},{\bf r}^\prime)$ describes the distribution of relative positions of the two dangling ends of a single-particle path that has been ``cut open".\\ Due to quantum exchanges (even as rare as they are in liquid hydrogen), and the ensuing entanglement of single-particle paths, $n(|{\bf r}-{\bf r}^\prime|)$ extends out to significantly greater distances than it would if particles were truly distinguishable (in which case $n(r)$ is very nearly Gaussian). This in turn results in a transfer of weight of $\bar n({\bf k})$ toward lower momenta, which can be interpreted as a sign of the incipient Bose-Einstein Condensation that liquid {\it para}-hydrogen would undergo, at a temperature around 5 K,  were it to escape crystallization. It has been suggested\cite{davidovsky} that this effect ought to be experimentally measurable.

The bulk of the available experimental information on the structure and dynamics of  liquid hydrogen comes from Deep Inelastic Neutron Scattering (DINS) measurements, which offer (in some cases direct) access to quantities such as the  single-particle kinetic energy, momentum distribution and excitation spectrum.\cite{lovesey} Such experiments have been pursued by various groups, over the past two decades.\cite{davidovsky,langel,Herwig,zoppi,zoppi2,milva0,milva,berme1,berme2} On the theoretical side, besides the above-mentioned numerical simulations, typically based on Quantum Monte Carlo (QMC) techniques, an analytical approach known as Correlated Density Matrix  (CDM) theory,\cite{davidovsky,qbl} has been shown to provide not only qualitative insight, but also quantitatively reliable predictions, at least for structural properties (an additional check on the predictions based on CDM is furnished here). 

Agreement between theoretical predictions and experimental data has not been of the quality that one would expect, given the availability of fairly well-established simulation methods and reasonably quantitative microscopic models. For example, the difference between the kinetic energy per particle in liquid \ph2 at $T$=16.5 K, in a range of pressure between 1 and 80 bars, computed theoretically\cite{qbl} and inferred from the most recent experimental measurements\cite{davidovsky} is approximately 10\% (a few K), which seems large, considering that the same comparison yields much better agreement for the more quantal helium liquid.\cite{azuah} It is unclear whether such a discrepancy originates within the microscopic model, or the computational methodologies, or may lie instead with the analysis of the experimental data. No results of any {\it first principles} microscopic calculation of the momentum distribution in liquid \ph2 have been reported so far.

The purpose of this paper is, on the one hand, to provide independent theoretical estimates for the kinetic energy per particle, as well as for the pair correlation function at the same conditions of temperature and pressure considered in Refs. \onlinecite{qbl} and \onlinecite{davidovsky}, thereby allowing for an  extended comparison between different calculations.  More importantly, an unbiased theoretical  estimate of the single-particle momentum distribution is furnished here, enabling a direct and cogent comparison between theory and experiment. The computational tool utilized here is numerical;  specifically, use is made of  the continuous-space Worm Algorithm (WA), which allows for a {\it direct} calculation of the one-particle density matrix, connected to the momentum distribution via a straightforward Fourier transformation.

Our numerical estimates for the kinetic energy are for the most part in quantitative agreement with those of other works, and thus retain the same level of disagreement with experimental data already reported by others. Quantum effects are clearly seen in the computed one-body density matrix, which can be directly associated to quantum exchanges, explicitly allowed in our calculation (i.e., {\it no} assumption of distinguishability is made here). In the next section, the microscopic model underlying the calculation as well as the basic features of the methodology utilized are reviewed. In Se. \ref{res} a thorough illustration of the results obtained in this work is provided; finally, a general discussion is offered, and conclusions outlined, in Sec. \ref{disc}.
\section{Model}
Consistently with all previous theoretical studies, our system of interest
is modeled as an ensemble of $N$ \ph2 molecules, regarded as point particles of spin zero,
enclosed in a cubic vessel of volume $\Omega$ with periodic boundary conditions.
The quantum-mechanical many-body Hamiltonian is the following:
\begin{equation}\label{one}
\hat H = -\lambda\sum_{i=1}^N \nabla_i^2 + \sum_{i<j} V(r_{ij}) 
\end{equation}
Here, $\lambda$=12.031 K\AA$^2$ for \ph2, while  
$V$ is the potential describing the interaction between two molecules, 
only depending on their relative distance. Most of the results presented here were obtained using the Silvera-Goldman model 
potential.\cite{SG} This is not the only potential that has been used in previous work, but it is arguably the
most commonly adopted. It
also been shown \cite{operetto} to afford an accurate quantitative description of the thermodynamics of the solid phase of \ph2. For the sake of comparison, however, we have also performed a calculation using the Buck\cite{buck1,buck2,colla}  pair potential. Naturally, in principle a more accurate model would go beyond the simple pair decomposition, including, for instance, also interactions among triplets; however, published numerical work (e.g., on helium) has given strong indications that three-body corrections, while significantly affecting the estimation of the pressure, have a relatively small effect on the structure and dynamics of the system, of interest here.\cite{bpc,syc}

The thermodynamic properties of the system, as modeled by the many-body Hamiltonian (\ref{one}), were studied by means of numerical simulations, based on the continuous-space Worm Algorithm (WA).\cite{worm,worm2} 
This (Monte Carlo) methodology, recently introduced, has several advantages over Path Integral Monte Carlo (PIMC), the technique utilized in most previous many-body numerical calculations for liquid \ph2. The most important, in this context, is the fact that it is formulated in an extended configuration space, including an open world line. This allows one to evaluate, besides all usual thermodynamic observables, also off-diagonal single-particle correlation functions not accessible in conventional PIMC, at no added computational cost. One of these correlation functions is the single-particle Matsubara Green function
\beq
g({\bf r},{\bf r}^\prime,\tau) = \langle {\hat {\cal T}} [\hat\psi({\bf r},\tau)\ \hat\psi^\dagger({\bf r}^\prime,0)]\rangle
\eeq
Here,  $\hat\psi$ and $\hat\psi^\dagger$ are time-dependent field operators, $-\beta\le \tau\le \beta$, with $\beta=1/T$ is commonly referred to as ``imaginary time",  and $\hat{\cal T}$ is the time ordering operator.\cite{pote} In the limit $\tau\to 0$, the Matsubara Green function reduces to the one-particle density matrix
$n({\bf r},{\bf r}^\prime)$, defined in Eq. (\ref{rel}).\\ For a translationally invariant system, it is $n({\bf r}^\prime,{\bf r})\equiv n({\bf r}^\prime-{\bf r})$, and in three dimensions the momentum distribution $\bar n({\bf k})$  is related to $n({\bf r})$ through
\beq\label{relm1}
n({\bf r})  = \frac{1}{(2\pi)^3}\ \int d^3k\ \bar n({\bf k})\ e^{i{\bf k}\cdot{\bf r}}
\eeq
For a system that is also isotropic, like a liquid, it is $n({\bf r})\equiv n(r)$. Hence, on inverting the above relationship one obtains
\beq\label{rel1}
\bar n({\bf k})  \equiv \bar n(k) = \frac{4\pi}{k}\  \int_0^\infty  dr\ r \ {\rm sin}{kr}\ n(r)
\eeq
For convenience, the following normalization is imposed on $\bar n(k)$:
\beq\label{norm}
\frac{1}{(2\pi)^3}\ \int d^3k\ \bar n({\bf k}) = 1
\eeq
which fixes to unity the value of the one-body density matrix at $r=0$. The average kinetic energy per particle $E_k$ is connected to the momentum distribution through
\begin{eqnarray}\label{kin}
E_k = \frac{1}{(2\pi)^3}\ \int d^3k\ \frac{\hbar^2 k^2}{2m}\ \bar n({\bf k})
\end{eqnarray}
Using (\ref{relm1}),  (\ref{norm}) and (\ref{kin}), it is straightforward to show that, in the limit $r\to 0$, it is
\beq
n(r) \approx 1 - \frac {E_k}{6\lambda}r^2
\eeq
which provides a useful consistency check on the computed $E_k$ and $n(r)$.
We obtain $\bar n(k)$ through (\ref{rel1}), using the numerically computed
$n(r)$.  

The reader is referred to Ref. \onlinecite{worm2} for a thorough description of  
the continuous-space WA.  The specific implementation 
utilized in this project is {\it canonical}, i.e., we keep the number $N$ 
of particles fixed.\cite{fabio}  Other technical aspects of the calculations 
are common to any other  QMC simulation scheme. Results were obtained by 
simulating systems comprising two different numbers of particles, namely
$N$=96 and $N$=256. Within the statistical errors of the calculations, no 
significant size dependence can be detected in any of the quantities considered here, except for the kinetic energy, for which a systematic difference of approximately 0.1 K exists for the two system sizes (the larger system yielding the greater value of $E_k$). \\
The usual fourth-order high-temperature propagator utilized in all previous
studies based on the WA was adopted here; convergence of 
the estimates was observed for a time step 
$\tau\approx 1.89\times 10^{-3}$ K$^{-1}$, which corresponds to 
$P$=32 imaginary time ``slices" at the single temperature considered 
in this work. 
Because the computational cost was negligible, all estimates reported 
here were obtained using twice as small a time step, in order to be on 
the safe side. 
\section{Results}
\label{res}
The thermodynamic conditions considered in this work are the same as in Ref. \onlinecite{qbl}. Specifically, the temperature is fixed at $T$=16.5 K, and three different densities are considered, namely $\rho$=0.02235 \Am3, 0.02372 \Am3 and 0.02413 \Am3, corresponding to pressures of 1, 40 and 80 bars respectively. 
\subsection{Structure}
\begin{figure}[h]
\centerline{\includegraphics[height=3.4in, angle=-90]{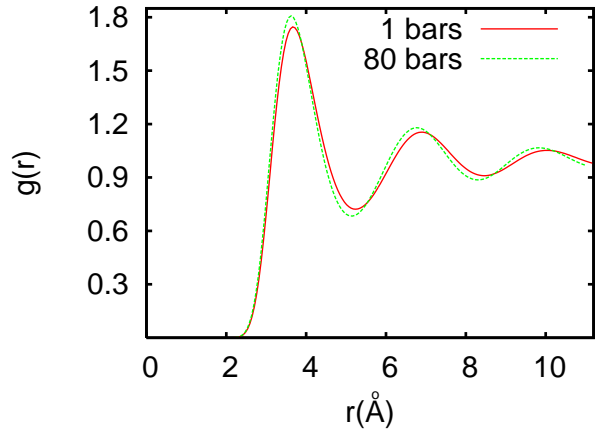}}
\caption{(Color online). Pair correlation function $g(r)$ in liquid \ph2 at $T$=16.5 K and $\rho$=0.02235 \Am3 (1 bar, solid line) and $\rho$=0.02413 \Am3 (80 bars, dashed line). Statistical errors are not visible, on the scale used here. }\label{gofr}
\end{figure}

Static correlation functions are yielded directly by the simulation. The results obtained here are largely consistent with those of previous MC calculations,\cite{znc,davidovsky} and shall therefore not be discussed any further.  The results at the lowest and highest densities considered here are shown in Fig. \ref{gofr}. The basic physical features of these functions have already been extensively described in the literature. Differences between different calculations, as well as between theory and experiment, if any, are not easily spotted.\cite{znc}

\subsection{Kinetic Energy}
\begin{table}[h]
\begin{tabular}{cccc} \hline
Density (\Am3)&0.02235 &0.02373 &0.02413 \\ \hline
Silvera-Goldman & 61.78(2) &64.99(5) &{67.93(4)}\\
Buck & 62.61(3) & & \\ \hline
PIMC (Ref. \onlinecite{qbl}) &62.7(4) &65.5(5)  &69.6(5) \\
PIMC (Ref. \onlinecite{milva}) &61.4(1) &  & \\
Expt. (Ref. \onlinecite{milva0}) &60.3(6) & & \\
Expt. (Ref. \onlinecite{davidovsky}) &67.8(3) & 73.5(4) &77.5(4)\\
\hline
\end{tabular}
\hfil\break 
\caption {\noindent {Kinetic energy per \ph2 molecule $E_k$ (in K) at $T$=16.5 K, computed in this work (first two rows) and in Refs. \onlinecite{qbl} and \onlinecite{milva}, as well as recently determined experimentally (Refs.  \onlinecite{milva0} and \onlinecite{davidovsky}). Statistical errors (in parentheses) are on the last digit. Theoretical estimates are obtained using the Silvera-Goldmann potential, except for that in the second row, which is based on the Buck potential. Experimental value from Ref. \onlinecite{milva0} is obtained as an interpolation of data at $T$=15.7 K ($\rho$=0.02252 \Am3) and $T$=17.2 K ($\rho$=0.02210 \Am3).}
}\label{tableone}
\end{table}
Table \ref{tableone} summarizes the kinetic energy per \ph2 molecule computed in this work; results quoted are for a system of $N$=256 molecules. For comparison, results of PIMC calculations published in Refs. \onlinecite{qbl} and \onlinecite{milva} are also shown, together with experimental data from Refs. \onlinecite{milva0}, as well as the more recent ones from Ref.  \onlinecite{davidovsky}. 
It should be noted that the kinetic energy in QMC is {\it not} computed through the momentum distribution, but rather using a direct estimator. \\
The comparison of theoretical estimates is not completely satisfactory, even after statistical and possible systematic uncertainties are properly taken into account. At the lowest density, the result offered in Ref. \onlinecite{milva} is very close to the one obtained here, and it is almost certain that the small difference is attributable to a relatively small number of time slices ($P$=64) used in Ref. \onlinecite{milva}, where the {\it primitive} approximation for the high-temperature propagator was utilized. While such a number is sufficient to obtain reasonably accurate estimate of structural properties (e.g., the pair correlation function), it was found in this work that at least twice as many slices are needed, in order to achieve convergence of kinetic energy estimates to the precision quoted here, if the primitive approximation is used (it is worth repeating that all results presented here are obtained using a fourth-order propagator).\\ 
On the other hand, the discrepancy between our estimates and those of  Ref. \onlinecite{qbl} is definitely outside statistical errors at the highest density, and possibly at the lowest one as well, in spite of the relatively large errors quoted therein; in general, the estimates offered in Ref.   \onlinecite{qbl} appear to be above ours, by almost 1.5-2 K at the highest density. Details of the calculation carried out in Ref. \onlinecite{qbl} are not available to us at this time, and therefore it is unclear what the source of such a disagreement might be, given that the model potential utilized is the same, and differences in system sizes seem unlikely to generate a disaccord of this magnitude.
\\
In order to carry out a comparison of the numerical results obtained here with those obtained by other methods, we have also performed a simulation at a density $\rho$=0.021 \Am3\ for a temperature $T$=16 K. For this thermodynamic condition, an estimate of the kinetic energy per particle of 58.6 K was provided in Ref. \onlinecite {davidovsky}, based on CDM theory. Our result  is 57.4 $\pm$ 0.1 K, which seems in reasonable agreement. 

A much more significant discrepancy exists between the theoretical and {\it some} of the available experimental estimates for the kinetic energy, namely those reported in Ref. \onlinecite{qbl}, all of which are above the theoretical ones, by approximately  6 K at the lowest density, up to almost 9 K at the highest. Such a disagreement is puzzling, considering the quantitative agreement between theoretically computed and experimentally measured kinetic energy per particle in liquid helium,\cite{azuah} in which deviations from the classical behavior are much more pronounced than in liquid hydrogen. In light of the substantial closeness (if not downright agreement) of the various independent theoretical estimates, one may look at the intermolecular potential as one of the possible sources of the discrepancy with experiment. The Silvera-Goldman potential has been shown to provide a rather accurate description of the crystalline phase of \ph2 at low temperature,\cite{operetto}  and therefore it seems unlikely that it would not be at least as adequate in the liquid phase, in which the consequences of the pair-wise spherical approximation should be even less important than in the crystal.

For comparison purposes, the same calculation was carried out in this work based on a different model potential as well, namely the Buck potential, at the lowest density considered here; the value is shown in Table \ref{tableone}. The Buck potential gives a slightly higher kinetic energy, by about 1 K, a considerably smaller difference than that between theory and experiment. It should be noted that both these potentials yield a theoretical value of the kinetic energy per particle in the solid phase within $\sim$ 1-2 K of the experimentally determined one.\cite{milva0,here} It seems therefore unlikely that the choice of pair potential may account for the observed difference between the theoretical estimates and the experimental data of Ref. \onlinecite{davidovsky}. 

Another possibility is that the finite size of the simulated system may result in an underestimation of the kinetic energy. However, as mentioned above calculations performed in this work on a system of $N$=96 particles, yield essentially the same estimate obtained with $N$=256, the difference being of the order of 0.1 K at the most. Therefore, it seems safe to exclude the possibility of a significant size dependence. Based on these considerations, we surmise that the discrepancy likely originates with some of the assumptions underlying the analysis of the experimental data carried out in Ref. \onlinecite{qbl}, which may have to be reconsidered. On this point, it is worth noting the substantial disaccord between the experimental estimates for the same quantity provided in Refs. \onlinecite{davidovsky} and \onlinecite{milva0} (about eight times the combined uncertainties quoted by the two groups). The value given in Ref. \onlinecite{milva0}, while not in perfect agreement with theoretical estimates, is nonetheless much closer to them than that furnished in Ref. \onlinecite{davidovsky}.

\subsection{Momentum Distribution}
\begin{figure}[h]
\centerline{\includegraphics[height=3.3in, angle=-90]{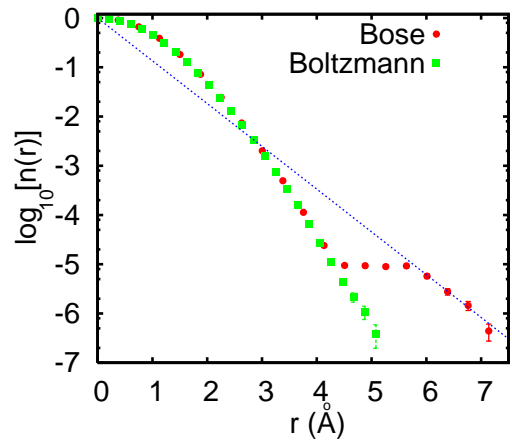}}
\caption{(Color online) Theoretically computed one-body density matrix $n(r)$ in liquid \ph2 at $T$=16.5 K and $\rho$=0.02235 \Am3 (circles). When not shown, statistical errors are smaller than symbol sizes. Straight line is an exponential fit to the curve for $r > 6$ \AA. Boxes show the result for $n(r)$  for {\it distinguishable} quantum particles; it is essentially identical with the one for Bosons for $r$ less than $\sim$ 4 \AA, but deviates significantly from it for larger distances, as a result of quantum exchanges. }\label{mom}
\end{figure}
Fig. \ref{mom} shows the computed (spherically averaged) one-body density matrix $n(r)$ for liquid \ph2 at the lowest density considered here, namely $\rho$=0.02235 \Am3.  Deviations from the classical (Gaussian) behavior are clear (at the same time, the logarithmic scale gives an idea of their magnitude).  In particular,  $n(r)$ extends beyond the first coordination shell, whose radius is $\sim$ 4 \AA, i.e., a molecule has a nonzero probability amplitude of exiting the ``cage"  formed by the surrounding molecules. This is due to quantum-mechanical exchanges, which have indeed been observed in all of the simulation carried out in this work (in the so-called $G$-sector -- see Ref. \onlinecite{worm} for details), and which allow the two dangling ends of the open path to drift further away from each other than if particles were distinguishable, by virtue of entanglement among different paths.

In order to establish this fact more quantitatively, we have performed a simulation in which particles were assumed truly distinguishable. In technical terms, this means that  the so-called ``swap" move,\cite{worm} which allows for entanglement of single-particle paths in the presence of a single open world line, is inhibited. The result that one obtains in this case for $n(r)$, also shown in Fig. \ref{mom},  is virtually {\it identical} with the one yielded by the calculation in which exchanges are included, for distances less than $\sim$ 4 \AA; for greater distances, on the other hand, it differs significantly, as the one-body density matrix continues to decay monotonically. On the scale of Fig. \ref{mom}, no signal appears for $r >$ 5 \AA, in such a simulation. This establishes that the structure of  $n(r)$ above 4 \AA\ is a genuine consequence of quantum (Bose) statistics.

From the numerically computed $n(r)$, one can obtain the experimentally observable momentum distribution $n(k)$ based on (\ref{rel1}), by means of a straightforward numerical integration. 
Obtaining good statistics for values of $n(r)$ at distances greater than $\sim$ 6 \AA\ becomes quickly impractical, due to the rapid decay of the function, which requires an exceedingly long simulation time in order to achieve a meaningfully small statistical uncertainty. This could in principle be an issue, when trying to evaluate (\ref{rel1}). However, the contribution to $n(k)$ coming from distances greater than 6 \AA\ can be easily estimated by fitting the portion of the curve for $r >$ 6 \AA, for example to an exponentially decaying function (as shown in Fig. \ref{mom}). As it turns out, such contribution (which is greatest at $k$=0) is barely worth 0.1\% of the total value of $n(k)$, regardless on the particular fitting function that one chooses, and therefore an accurate evaluation of (\ref{rel1}) can be obtained by integrating only up to $r$=6 \AA.

\begin{figure}[h]
\centerline{\includegraphics[height=3.4in, angle=-90]{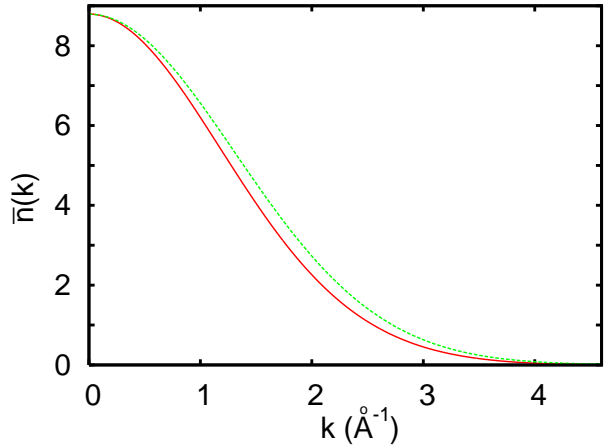}}
\caption{(Color online). Momentum distribution $\bar n(k)$ in liquid \ph2 at $T$=16.5 K and $\rho$=0.02235 \Am3. Solid line is the estimate obtained by numerical integration of Eq. \ref{rel1}, using the results for $n(r)$ shown in Fig. \ref{mom}, as explained in the text. Statistical and systematic errors are too small to show on the scale of the figure. Dashed line is a Gaussian function whose width is chosen so as to yield the kinetic energy per particle computed in the simulation, namely 61.78$\pm$0.02 K. }\label{mom2}
\end{figure}
\begin{figure}[h]
\centerline{\includegraphics[height=3.4in, angle=-90]{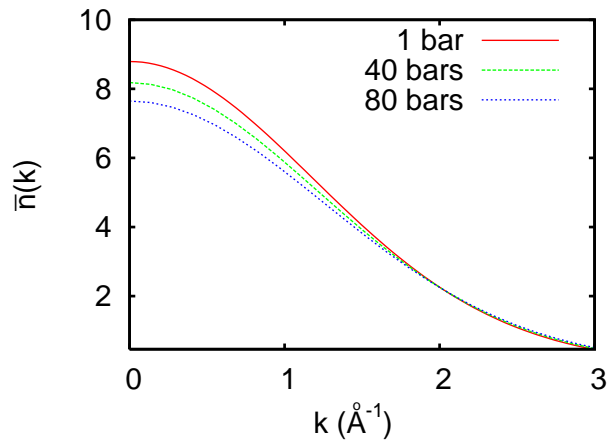}}
\caption{(Color online). Momentum distribution $\bar n(k)$ in liquid \ph2 at $T$=16.5 K and $\rho$=0.02235 \Am3 (1 bar, solid line), $\rho$=0.02327 \Am3 (40 bars, dashed line), and $\rho$=0.02413 \Am3 (80 bars, dotted line).  Statistical errors are not visible, on the scale used here. }\label{cmp}
\end{figure}

Fig. \ref{mom2} shows the momentum distribution $\bar n(k)$ resulting from the numerical integration of (\ref{rel1}) based on the data for $n(r)$ shown in Fig. \ref{mom}. Although the system studied is not a classical liquid, the overall shape of $\bar n(k)$ remains close to a Gaussian,\cite{withers} albeit  one corresponding to a different, {\it effective} temperature, given by 2$E_k$/3 (shown for comparison in Fig. \ref{mom}).  Such a Gaussian is essentially indistinguishable from the momentum distribution that one obtains on Fourier transforming the one-body density matrix computed for distinguishable quantum particles.
There is some strength transferred to both lower and higher momenta, compared to what one would find if the momentum distribution were indeed such a ''renormalized" Gaussian; more quantitatively, numerical integration of Eq. (\ref{kin}) up to momentum $k_\circ$=4 \AA$^{-1}$, using the data for $\bar n(k)$ shown in Fig. \ref{mom2}, yields approximately 85\% of the overall kinetic energy (as opposed to 92.8\% if $\bar n(k)$ were a Gaussian), the rest coming from momenta higher than $k_\circ$. This underscores the delicacy of extracting the single-particle kinetic energy from the experimentally measured momentum distribution, as an accurate determination of the tail is required, a fact which might help account some of the differences reported in the literature between the numerically computed and experimentally determined kinetic energy.\cite{diallo}

\begin{figure}[h]
\centerline{\includegraphics[height=3.3in, angle=-90]{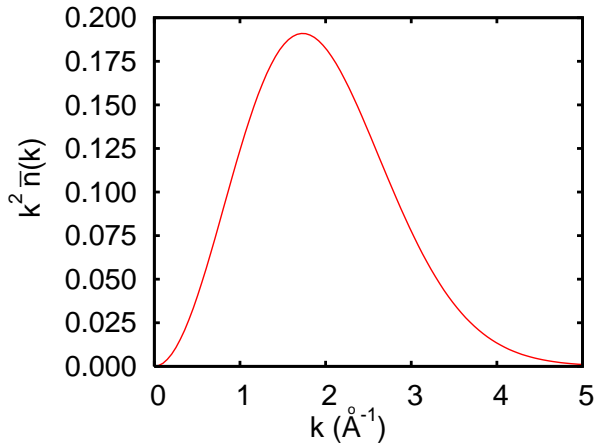}}
\caption{(Color online). Kinetic energy distribution $k^2 \bar n(k)$ in liquid \ph2 at $T$=16.5 K and $\rho$=0.02235 \Am3 (1 bar, solid line). Statistical errors are not visible, on the scale used here. The same units are used as in Ref. \onlinecite{davidovsky}. }\label{caz}
\end{figure}

The difference between the computed momentum distribution and the model Gaussian, which can be attributed entirely to quantum statistics, is altogether rather small, and is most noticeable for momenta between 1.5 and 2.5 \AA$^{-1}$, corresponding to interparticle distances between 2.5 and 4 \AA. This conclusion was already stated in Ref. \onlinecite{davidovsky}, based on an analysis of experimental data for $\bar n(k)$. In Fig. \ref{caz} the kinetic energy distribution is shown, namely the quantity $k^2 \bar n(k)$, for which experimental data are  reported in Ref. \onlinecite{davidovsky} (the same units utilized therein are used). Within the statistical uncertainties of the calculation carried out here, there appears to be broad agreement between the results obtained in this work and the experimental data, but with some noticeable differences. In particular, the computed $k^2 \bar n(k)$ attains its maximum in correspondence of the momentum $k$ = 1.735$\pm$0.005 \AA$^{-1}$, in excellent agreement with the value 1.736 \AA$^{-1}$ reported in Ref. \onlinecite{davidovsky}. On the other hand, around $k\approx$ 4 \AA$^{-1}$ the curve calculated here falls below the experimental one, approximately by a factor 2.5. This is consistent with the fact that the experimentally determined kinetic energy is above that obtained here by approximately 10\%.

The momentum distribution at higher density (pressure) does not qualitatively change, with respect to that shown in  Fig. \ref{mom}. This is shown in  Fig. \ref{cmp}, where the momentum distribution computed at the three different densities considered here is displayed. The most important change that occurs on raising the pressure is the loss of weight at low momenta, as the system becomes increasingly classical. Correspondingly, the tail of the one-body density matrix at long distances is suppressed, as exchanges are rarer and rarer, even if one single-particle world line is open. 
\bigskip

\section{Conclusions}\label{disc}
We have carried out first principle calculations of the momentum distribution of liquid {\it para}-hydrogen at $T$=16.5 K, at three different densities corresponding to pressures ranging between 1 and 80 bars.  The results for the one-particle density matrix show that quantum-mechanical exchanges result in a longer tail of the one-body density matrix than one would predict based on distinguishability of molecules. Consequently, this has an effect on the momentum distribution, which features clear, measurable deviations from a Gaussian. These conclusions are in broad qualitative and quantitative agreement with considerations made in Ref. \onlinecite{davidovsky}. On increasing the pressure, exchanges are suppressed, and the most significant change in the momentum distribution occurs precisely at low $k$.

For the thermodynamic conditions explored here, experimental measurements have been recently carried out. The agreement between theoretical estimates and experimental data continue to be less than satisfactory, certainly of much lesser quality than that found for liquid and solid helium. The deviation between theoretical and experimental data appears to be systematic.
At the time of this writing, it is unclear to us where the origin of the disagreement lies, but perhaps an independent check of the analysis of the data of Ref. \onlinecite{davidovsky} is in order.

As mentioned above, the discrepancy could also be attributed to an inaccurate determination of the tail of the momentum distribution. However, another aspect worth revisiting is the effect of multiple inelastic scattering that a neutron suffers in a liquid. An example of the impact of multiple scattering is discussed in Ref. \onlinecite{diallo2}, where it is shown that it can indeed lead to an overestimation of the value of the center-of-mass kinetic energy per particle. 

\section*{Acknowledgments}

This work was supported in part by the Natural Science and Engineering Research Council of Canada under research grant 121210893, and by the Alberta Informatics Circle of Research Excellence (iCore).
The author gratefully acknowledges the hospitality of the Institut f\"ur Theoretische Physik, Universit\"at Innsbruck, Austria, as well as illuminating discussions with Milva Celli, Daniele Colognesi, Henry Glyde and Nikolay Prokof'ev.


\begin{thebibliography}{99}

\bibitem{intro} Henceforth, when talking about the kinetic energy of hydrogen molecules we shall always refer to the {\it center of mass} contribution alone.

\bibitem{note0} We adopt here a system of units where the Boltzmann constant $k_B$=1.

\bibitem{feynman}
See, for instance, R. P. Feynman, {\it Statistical Mechanics: A Set of Lectures}, (Addison-Wesley, New York, 1972).
\bibitem{qbl} See, for instance, K. A. Gernoth, T. Lindenau and M. L. Ristig, Phys. Rev. B {\bf 75}, 174204 (2007).
\bibitem{neumann91}
M. Neumann and M. Zoppi, Phys. Rev. A {\bf 44}, 2474 (1991).
\bibitem{zoppib}
M. Zoppi and R. Neumann,  Physica B {\bf 180-181}, 825 (1992).
\bibitem{wang97}
Q. Wang, J. K. Johnson and J. Broughton, J. Chem. Phys. {\bf 107}, 5108 (1997).
\bibitem{kin}
K. Kinugawa,  Chem. Phys. Lett. {\bf 292}, 454 (1998).
\bibitem{znc}
M. Zoppi, M. Neumann and M. Celli, Phys. Rev. B {\bf 65}, 092204 (2002).
\bibitem{milva2}
M. Celli, U. Bafile, G. J. Cuello, F. Formisano, R. Magli, E. Guarini, M. Neumann and M. Zoppi, Phys. Rev. B {\bf 71}, 014205 (2005).
\bibitem{note}
In the crystalline phase, quantum exchanges are suppressed by particle localization, much as in helium. See, for instance, M. Boninsegni, N. Prokof'ev and B. Svistunov, Phys. Rev. Lett. {\bf 96}, 105301 (2006).
\bibitem{davidovsky}
 J. Dawidowski, F. J. Bermejo, M. L. Ristig, C. Cabrillo, and S. M. 
Bennington, Phys. Rev. B {\bf 73}, 144203 (2006).

\bibitem{lovesey}
S. W. Lovesey, {\it Theory of Neutron Scattering from Condensed Matter}, (Clarendon press, Oxford 1987).

\bibitem{langel}
W. Langel, D. L. Price, R. O. Simmons, P. E. Sokol, Phys. Rev. B {\bf 38}, 11275 (1988).
\bibitem{Herwig}
K. W. Herwig, J. L. Gavilano, M. C. Schmidt, R. O. Simmons, Phys. Rev. B {\bf 41}, 96 (1990).
\bibitem{zoppi}
 M. Zoppi, U. Bafile, R. Magli, and A. K. Soper, Phys. Rev. E {\bf 48}, 1000 
(1993) 
\bibitem{zoppi2}
M. Zoppi, U. Bafile, E. Guarini, F. Barocchi, R. Magli, and M. Neumann, Phys. Rev. Lett. {\bf 75}, 1779 (1995).
\bibitem{milva0}
M. Celli, D. Colognesi, and M. Zoppi, Eur. Phys. J. B {\bf 14}, 239 (2000).
\bibitem{milva}
M. Celli, D. Colognesi, and M. Zoppi, Phys. Rev. E {\bf 66}, 021202 (2002).
\bibitem{here}
A MC calculation performed in this work for solid \ph2 at $T$=12.2 K and $\rho=0.02602$ \Am3, based on the Silvera-Goldman potential, yields a value of $E_k$ for the solid of 70.9(1) K, to be compared with the PIMC estimate of 70 K (no statistical uncertainty is quoted)  and the experimental result of 68(2) K offered in Ref. \onlinecite{milva0} for the same thermodynamic conditions.
\bibitem{berme1}
F. J. Bermejo, B. F\aa k, S. M. Bennington, K. Kinugawa, J. Dawidowski, 
M. T. Fern\'andez-D\'iaz, C. Cabrillo and R. Fern\'andez-Perea, Phys. Rev. B {\bf 66}, 212202 (2002).
\bibitem{berme2} 
J. Dawidowski, F. J. Bermejo, M. L. Ristig, B. F\aa k, 
C. Cabrillo, R. Fern\'andez-Perea, K. Kinugawa, and J. Campo, Phys. Rev. B. {\bf 69}, 014207 (2004). 

\bibitem{azuah}
See, for instance, R. T. Azuah, W. G. Stirling, H. R. Glyde, M. Boninsegni, P. E. Sokol and S. M. Bennington, Phys. Rev. B {\bf 56}, 14620 (1997).
\bibitem{SG}
I. F. Silvera and V. V. Goldman, J. Chem. Phys. {\bf 69}, 4209 (1978).
\bibitem{operetto}
F. Operetto and F. Pederiva,  Phys. Rev. B {\bf 73}, 184124 (2006).
\bibitem{buck1} U. Buck, F. Huisken, A. Kohlhase, D. Otten, and J. Schaeffer
J. Chem. Phys. {\bf 78}, 4439 (1983).
\bibitem{buck2} M. J. Norman, R. O. Watts and U. Buck, J. Chem. Phys.
{\bf 81}, 3500 (1984). 
\bibitem{colla}
For a comparison of the Silvera-Goldman and Buck potentials,
see, for instance, E. Cheng and K. B. Whaley, J.
Chem. Phys. {\bf 104}, 3155 (1996).
\bibitem{bpc}
M. Boninsegni, C. Pierleoni and D. M. Ceperley, Phys. Rev. Lett. {\bf 72}, 1854 (1994).
\bibitem{syc}
S.-Y. Chang and M. Boninsegni, J. Chem. Phys. {\bf 115}, 2629 (2001).
\bibitem{worm}
M. Boninsegni, N. Prokof'ev and B. Svistunov, Phys. Rev. Lett. {\bf 96}, 070601 (2006).
\bibitem{worm2}
M. Boninsegni, N. Prokof'ev and B. Svistunov, Phys. Rev. E {\bf 74}, 036701 (2006).
\bibitem{pote} See, for instance, A. Fetter and J. D. Walecka, {\it Quantum Theory of Many-particle Systems} (McGraw-Hill, New York, 1973).
\bibitem{fabio}
F. Mezzacapo and M. Boninsegni, Phys. Rev. Lett. {\bf 97}, 045301 (2006).
\bibitem{withers}
See, for instance, B. Withers and H. R. Glyde, J. Low Temp. Phys. {\bf 147}, 633 (2007).
\bibitem{diallo}
See also, for instance, S. O. Diallo, J. V. Pearce, R. T. Azuah, F. Albergamo and H. R. Glyde, Phys. Rev. B {\bf 74}, 144503 (2006).
\bibitem{diallo2}
S. O. Diallo, J. V. Pearce, R. T. Azuah, O. Kirichek, J. W. Taylor and H. R. Glyde, Phys. Rev. Lett. {\bf 98}, 205301 (2007).
\end{thebibliography}
\end{document}